# On the relationship between velocities, tractions, and intercellular stresses in the migrating epithelial monolayer

Yoav Green[1,2*], Jeffrey J. Fredberg[1], and James P. Butler[1,3]

[1]*Harvard T.H. Chan School of Public Health, Boston, MA 02115, USA.*
[2] *Department of Mechanical Engineering, Ben-Gurion University of the Negev, Beer-Sheva 8410501, Israel*
[3] *Department of Medicine, Brigham and Women's Hospital, Harvard Medical School, Boston, MA 02115, USA.*
\* Correspondence: yoavgreen@bgu.ac.il

The relationship between velocities, tractions, and intercellular stresses in the migrating epithelial monolayer are currently unknown. Ten years ago, a method known as Monolayer Stress Microscopy (MSM) was suggested from which the intercellular stresses could be computed given a traction field. The core assumption of MSM is that the intercellular stresses within the monolayer behave similarly to passive systems like a Hookean solid (an elastic sheet) or a Newtonian fluid (thin fluid film), implying a relation between the displacements/velocities and tractions. Due to the lack of independently measured intercellular stresses, validation of MSM is difficult. An alternative approach, which we give here, is based on simultaneous measurements of the monolayer velocity field and the cell/substrate tractions. With limited assumptions, the velocity field suffices to compute tractions, which we can then compare directly with those measured by traction force microscopy. We find that the calculated tractions and measured tractions are uncorrelated. Since both classical MSM and a purely viscous description of the relation between displacements or velocities and tractions depends on a linear constitutive law, it follows that some modification of these approaches is needed. One possible resolution is the inclusion of an active force. To this end, we give a new relationship between the active force density and the measured velocity (or displacement) field and tractions, which by Newton's laws, must be obeyed.

## I. INTRODUCTION.

During wound healing, cancer metastasis, development, and asthmatic airway remodeling, cells comprising a confluent epithelial layer migrate collectively [1–5]. Within the epithelial monolayer, each constituent cell exerts intercellular stresses on neighboring cells, and exerts traction forces [6] on its substrate. While traction forces exerted by a monolayer have been measured for two decades [7–9] , their relationship to measured cellular velocities remains unknown, and the relationship between intercellular stresses and tractions remains unresolved [10].

Ten years ago [11,12], our group suggested a method to recover the induced stresses, $\sigma$, within a monolayer given a measured traction vector field, $T$. The method was termed Monolayer Stress Microscopy (MSM) [11,12] and we denote the recovered stresses as $\sigma_{MSM}(T)$. The principle assumption was that the stresses could be described by the same equations describing simple passive systems such as Hookean solids [13] or Newtonian fluids [14,15], where there is a linear relationship between stress and strain [13], or stress and strain rate, respectively [15]. Since our first work, [11], this approach has been adopted by many, and various derivative formulations have been suggested (Hookean solid [10–12,16–28] or Newtonian fluid [10,18,20,29–31], see recent review [10] for more details).

It is clear that this approach is exact in 1D. Because where the divergence of the stress is given by the traction, no assumptions are required regarding the constitutive law in 1D. In 2D, however, the situation is different, and a constitutive law is required. Additional difficulties can be attributed to the effects of the boundary conditions (described below). Tambe et al. [12]

addressed the Poisson ratio issue by computing intercellular stresses with a wide variety of Poisson ratios. They found the dependence upon the Poisson ratio to be small, thus lending credence to the MSM approach. On the other hand, it has also been suggested that under certain conditions, including a Poisson ratio of zero, the 2D problem decouples into two 1D (exact) problems. This statement is incorrect.

Nevertheless, MSM remains to be validated in the following sense: the tractions are the independent variable, along with boundary conditions at the edge of the monolayer or field of view (see Section II below) and then $\sigma_{MSM}(T)$ is calculated. But we currently have no direct measures of intercellular stress by which to compare these. An equivalent test validating the use of a passive constitutive law is to note that a measured local velocity or displacement field, together with an assumed viscosity or shear modulus, suffices to compute the tractions. And these in turn can then be compared with the experimentally measured tractions using traction force microscopy (TFM) [7–9]. This comparison is the focus of this work. We find that tractions computed from the measured velocity field are uncorrelated with tractions measured by TFM.

This paper is structured as follows. In Sec. II we review MSM used for stress recovery within a monolayer [11]. Since its conception ten years ago [11], many derivative models have been suggested that use either a Hookean solid [10–12,16–28] or a Newtonian fluid [10,18,20,29–31] constitutive equation. We review both formulations and most embedded assumptions of MSM. In Sec. III, we present our alternative approach and show that MSM is an incomplete theory. Sec. IV discusses the implications of our result, which is that the constitutive equation needs modification. One of the suggested modifications is the inclusion of active stresses. A variety of such active stresses formulations have been put forward on an *ad hoc* basis [10,17,19–21,26–31]. By contrast with *ad hoc* formulations, here we put forward a fundamental relationship that links active force density to substrate tractions and velocity fields, and is derived with a minimal amount of assumptions. Because it follows directly from Newton's laws and does not depend on a specific biophysical rheology or molecular mechanism, this relationship is general and robust. Finally, we conclude in Sec. V with a few short remarks.

## II. MONOLAYER STRESS MICROSCOPY

In this section, we review MSM as suggested by us and its implementation by others with various derivative models. Sec. II.A presents the general three dimensional (3D) governing equations. Sec. II.B discusses constitutive equations while Sec. II.C discusses incompressibility. In Sec. II.D, we reduce the governing equation to two dimension (2D). Sec. II.E discusses the various boundary conditions (BCs) needed at the edge of the monolayer.

### A. 3D model

In a 3D bulk where inertia is neglected and there are no body forces [Figure 1(a)], Newtons's second law is

$$\nabla_{3D} \cdot \boldsymbol{\sigma}_{3D} = 0, \tag{1}$$

where $\boldsymbol{\sigma}_{3D}$ are the 3D stresses. In 3D, the 2D tractions, $\boldsymbol{T}$, are accounted for as BCs [Figure 1(b)] while BCs at the edge of the monolayer are discussed in Sec. II.E. In the remainder of this work, we will refer to $\boldsymbol{T}$ as those measured by traction force microscopy – TFM [8,9] (i.e. the forces the cells apply on the substrate). Eq. (1) is an indeterminate vector of three equations comprising six different terms $(\sigma_{ij}, i, j = x, y, z)$. Hence, solution of this equation requires additional equations in the form of either a constitutive equation or compatibility equation.

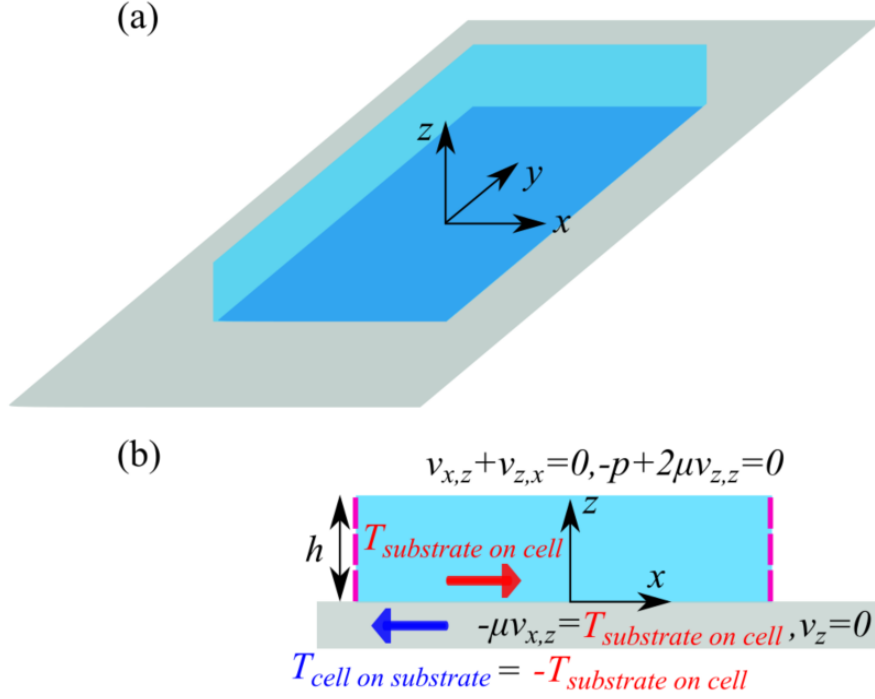

Figure 1. (a) 3D schematic of a thin film of fluid (or epithelial monolayer) on a rigid substrate. Not drawn to scale – the length and width of the film is substantially larger than the height. (b) $x-z$ cross-section. The top surface is stress free while at the bottom, tractions are balanced by viscous shear stresses. In this work, tractions ($T_x$ and $T_y$) refer to the measured tractions by the cells on the substrate, $T_{cell\ on\ substrate}$. The dashed lines refer to boundary conditions at the edges discussed in Sec. II.E.

**B.    Elastic and viscous constitutive laws**

The original MSM model [11,12] was formulated in 2D. There, the 2D equations comprised three terms ($\sigma_{xx}, \sigma_{yy}, \sigma_{xy}$). To close these set of equations, the Beltrami-Mitchell compatibility equation was supplemented [11,12]. However, using this equations is tantamount to assuming that the monolayer behaves as a simple Hookean constitutive law [13] characterized by two material constants: shear modulus, $G$, and Poisson ratio, $\nu$. In this work, for the sake of simplicity, we consider the case of $\nu = \frac{1}{2}$, which is the limiting case of an incompressible elastic material (i.e. volume conservation – see Sec. II.C). For an incompressible elastic medium the constitutive law [13]

$$\boldsymbol{\sigma} = \tfrac{1}{3} tr(\boldsymbol{\sigma})\boldsymbol{I} + G(\boldsymbol{\nabla}_{3D}\boldsymbol{u}_{3D} + \boldsymbol{\nabla}_{3D}\boldsymbol{u}_{3D}^{T}), \qquad (2)$$

where $\boldsymbol{u}_{3D}$ is the 3D displacement vector field. However, without time-dependent terms, in Eq. (1) the equations for an incompressible Hookean solid and incompressible Newtonian fluid are identical, except for a time derivative in the latter case. This is easily seen by inspection of the constitutive equation for an incompressible fluid [14,15]

$$\boldsymbol{\sigma} = \tfrac{1}{3} tr(\boldsymbol{\sigma})\boldsymbol{I} + \mu(\boldsymbol{\nabla}_{3D}\boldsymbol{v}_{3D} + \boldsymbol{\nabla}_{3D}\boldsymbol{v}_{3D}^{T}), \qquad (3)$$

where $\boldsymbol{v}_{3D}$ and $\mu$ are respectively the 3D velocity vector field and viscosity. It is immediately clear that Eqs. (2) and (3) are identical with the identification of displacement with velocity, and with shear modulus and viscosity. How can these two distinctly different physical models give the same prediction? From the experimental perspective, what is typically measured is the displacement of cells, $\boldsymbol{u}$, between two consecutive frames, say at $t=0$ and $t=\Delta t$. Importantly, if the stress field at $t=0$ is assumed to be zero, then whether one uses the displacement field, Eq. (2), or the velocity field, Eq. (3), (from $\boldsymbol{v} = \boldsymbol{u}/\Delta t$) the stresses from both approaches are necessarily the same. This addresses the issue of whether a Hookean description can be valid when cells are clearly moving. The answer is yes insofar as for any pair of images, the assumption a stress-free state in the first image yields the exact same stress field as the Newtonian fluid description. Importantly, the fluid description is more general in that does not require an assumption of a zero stress state in any of the image frames, nor does it require an assumption of small strains.

Thus both models are essentially identical, here we adopt the viscous formulation as it requires fewer assumptions from the experimental perspective. In what follows, we will use the more conventional notation for the fluid's constitutive equation

$$\sigma = -p\mathbf{I} + \mu(\nabla_{3D}\mathbf{v}_{3D} + \nabla_{3D}\mathbf{v}_{3D}^T),\qquad(4)$$

whereby the pressure, $p$, takes the place of the trace of the stress tensor, $p \triangleq -\frac{1}{3}tr(\sigma)$. Regardless of the chosen experimental interpretation, since the models are identical, both formulations are equivalent and share the discrepancy that will be discussed in Sec.III.

Inserting any of these constitutive equations [Eq. (2)-(4)] into Eq. (1) leads to

$$-\nabla p + \mu\nabla^2 \mathbf{v}_{3D} = 0.\qquad(5)$$

### C. Incompressibility

Cells are comprised mainly of water and as such are essentially incompressible. In either the Hookean solid formulation [13] or Newtonian fluid formulation [14,15], an incompressible material must also satisfy volume conservation. In 3D, this reads as

$$\nabla_{3D} \cdot \mathbf{v}_{3D} = 0.\qquad(6)$$

Nevertheless, cells are not isovolumic insofar water can be transported into or out of the cell via osmotic stress or water channels. Intercellular fluid flow within a monolayer occurs over timescales of hours [32–34]. These water fluxes comprise a source term in the continuity equation [Eq. (6)]. Indeed, it has been shown that osmotic pressure results in cell area and volume oscillations with a period of four hours and amplitude of 20% [32,33]. Further, proliferation and apoptosis, and cell extrusion form the layer represent additional source terms. [26,35]. Because these phenomena are slow (~hours) compared with the time scales of interest here ($\Delta t$ ~minutes), we take the monolayer to be isovolumic, as have others [10,19,20,24,36–40].

### D. Reduction from 3D to 2D

In the case of plane stress, the 3D equations for either elastic bodies or fluids can be reduced to thin fluid film in 2D [13]. The reduction is based on the assumption that the height of the monolayer, $h$, is much smaller than characteristic in-plane lengths and is taken to be uniform [9,10,18].

Here we give a brief derivation in the incompressible case (see chapter 67 of Sokolnikoff [13]). In plane stress, the shear stresses are zero

$$\sigma_{xz} = \sigma_{yz} = 0, \tag{7}$$

and the $z$ component of the normal stress is zero ($\sigma_{zz} = 0$). The constitutive law [Eq.(4)] yields

$$\sigma_{zz} = -p + 2\mu v_{z,z} = 0 \Rightarrow v_{z,z} = p/(2\mu). \tag{8}$$

The comma subscript denotes partial differentiation. From 3D incompressibility [Eq.(6)] we have

$$\nabla_{3D} \cdot \mathbf{v}_{3D} = 0 \Rightarrow v_{z,z} = -v_{x,x} - v_{y,y} = -\nabla_2 \cdot \mathbf{v}_{2D}, \tag{9}$$

where $\nabla_2 \equiv (\partial_x, \partial_y)$ is a 2D operator and $\mathbf{v}_{2D} = (v_x, v_y)^T$. Equating Eqs. (8)-(9) yields the 2D incompressibility condition

$$\nabla_2 \cdot \mathbf{v}_{2D} = -p/(2\mu). \tag{10}$$

In contrast to a 3D formulation, where the tractions are imposed as boundary conditions [41], in the plane stress reduction the tractions assume the role of a body force [10–12,16–24,29]. It is common to write the 2D governing equation of force balance as

$$\nabla_{2D} \cdot \boldsymbol{\sigma}_{2D} = \mathbf{T}/h. \tag{11}$$

However, it is beneficial to use the constitutive equation [Eq.(4)]. Then, this reads as

$$\sigma_{ij,j} = -p_{,i} + \mu(v_{i,jj} + v_{j,ji}) = T_i/h, \tag{12}$$

Using Eq. (10), Eq. (12) becomes

$$\sigma_{ij,j} = -\tfrac{3}{2} p_{,i} + \mu v_{i,jj} = -T_i/h, \tag{13}$$

or, alternatively, in vector notation

$$-\tfrac{3}{2} \nabla_{2D} p + \mu \nabla_{2D}^2 \mathbf{v}_{2D} = \mathbf{T}/h, \tag{14}$$

Typically, MSM is presented by Eq. (11) along with the incompressible Beltrami-Mitchell equation

$$\nabla^2_{2D}(\sigma_{xx} + \sigma_{yy}) = \tfrac{3}{2} \boldsymbol{T}/h. \tag{15}$$

We call these two equations [Eqs. (11) and (15)] the classical MSM formulation, in which the tractions are the independent variable and the stresses are the dependent variable. By contrast, in Eqs. (10) and (14), the tractions are again the independent variable, but now the velocities are the dependent variable, from which the stresses can be calculated. We call these two equations the Stokes formulation. As noted in Sec. II.B it is important to recognize that both formulations are identical. If all the BCs at the edge of the monolayer be given in term of the stresses, then Eqs. (11) and (15) can be solved. However, as we will discuss in the next sub-section, typically the BCs are given in term of the velocities, and as such, Eqs. (11) and (15) cannot be solved without the introduction of an intermediate variable of displacement or velocity through the constitutive law. In strong contrast, Eqs. (10) and (14) can be solved directly and locally for tractions with velocities taken as the independent variable without the need for specific BC assumptions. This will be used below in Sec. III, where the tractions obtained from Eqs. (10) and (14) are used to directly compare with experimentally measured tractions.

In the remainder of this work, where we consider the 2D MSM problem and for the sake of brevity, we drop the 2D subscript.

**E.     Edge boundary conditions**

Originally, MSM [11,12] estimated intercellular stresses directly from the traction field. Solution of the governing equations requires assumptions about the boundary conditions at the free edges of the cellular domain or at the boundary of the field of view. This latter boundary was termed the "optical edge" [12]. In what follows we will refer to the conditions at both the free and optical edges as "edge conditions", to distinguish these from the 3D BCs at the top and bottom surfaces discussed in Figure 1.

Here we revisit the question of what are the most appropriate edge conditions necessary for a computation of intercellular stresses. We begin with a short review of our previous discussion [12]. Figure 2 shows a schematic representation of an island of cells expanding into free space where four scenarios are considered. The island's boundary is given by a solid black line. The red circle represents an obstacle. Each square represents a possible field of view, and is color coded for four possible cases. In all cases, we assume velocities (or displacements) and tractions are accessible through measurement.

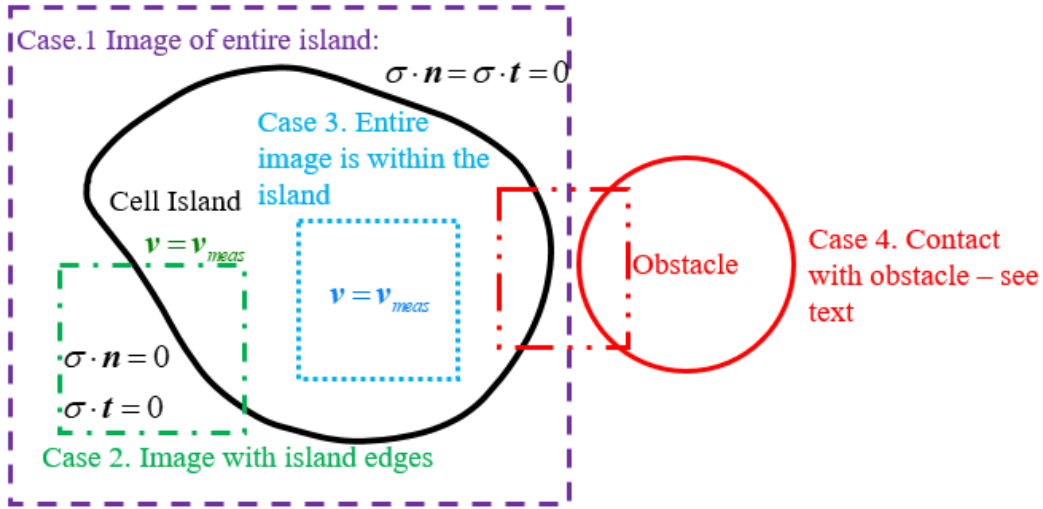

Figure 2. Schematic of a cell island approaching an obstacle. The cell island is given by a solid black line. Different scenarios, where the cell island is imaged fully or partially, are color coded. See text for case 4 edge conditions.

Case 1 (dashed purple line). The entire island is imaged. In this situation, the edge conditions at the free edges are stress free:

$$\boldsymbol{\sigma} \cdot \boldsymbol{n} = 0, \boldsymbol{\sigma} \cdot \boldsymbol{t} = 0. \tag{16}$$

Case 2 (dashed-dotted green line). Only part of the island is imaged. For the free edges, the edge conditions are given by Eq. (16), as in Case 1. However, for the optical edges (at the boundaries of the field of view) an assumption must be made. Previously [12], we argued that at the "optical" boundaries the condition is zero normal flux and zero shear stress

$$\boldsymbol{v} \cdot \boldsymbol{n} = 0, \boldsymbol{\sigma} \cdot \boldsymbol{t} = 0. \qquad (17)$$

Here we suggest that, instead of this ad hoc assumption, it be replaced by using additional measurements of velocities or displacements at the optical edge (e.g. with PIV) such that at the optical edge,

$$\boldsymbol{v} = \boldsymbol{v}_{meas}. \qquad (18)$$

Other approaches are possible, including requiring zero tractions at the cell free boundary [23,24], with an apparent stress jump at the same boundary, the origins of which remain open.

Case 3 (dotted blue line) is solely determined by optical edges, for which Eq. (18) is the appropriate edge condition.

Case 4 (long-dashed dotted red line in Figure 2), when cells have contacted an obstacle. Here we must make a distinction between two kinds of obstacles, each of which as two limiting subtypes. The first is a physical obstacle. In this case, the first subtype is no-slip and no flux, for which, at the interface,

$$\boldsymbol{v} = 0. \qquad (19)$$

The second subtype is a pure slip (i.e. zero shear stress) with no flux, for which, again at the interface, the edge condition is

$$\boldsymbol{\sigma} \cdot \boldsymbol{t} = 0, \boldsymbol{v} \cdot \boldsymbol{n} = 0. \qquad (20)$$

The other kind of obstacle is chemical, for example, the boundary between a region coated with collagen and region without. Here, the first subtype is where there are no cellular interactions outside the observed boundary of the island, for which the edge condition is stress-free [Eq. (16)]. The second subtype is where there may be cellular interactions and non-zero tractions outside the visible island. For example, the presence of protruding lamellipodia may exert tractions but may not be visible on a phase image. In this case, it is most appropriate to use zero normal velocity and zero stress [Eq.(20)]. Which condition is most befitting depends on the nature of the obstacle and demands attention to the biology of cell/ECM interactions.

Finally, we note the approach of two recent works [30,31]. These works differ from this one, whose starting point is a purely passive constitutive equation, in that their initial assumption was that the stress tensor included an active term [similar what will be discussed in Sec. IV]. The inclusion of their stress term added an additional free parameter that could be determined by applying a tangential and two normal BCs at the free edge

$$\boldsymbol{\sigma} \cdot \boldsymbol{t} = 0, \boldsymbol{\sigma} \cdot \boldsymbol{n} = 0, \boldsymbol{v} \cdot \boldsymbol{n} = 0 .\tag{21}$$

It should be noted that this approach worked in their quasi 1D geometry of a circular island. However, it remains to be seen if a single fitting parameter is suitable for arbitrary 2D geometries.

### III. MSM VALIDATION

This section is divided as follows. Sec. III.A discusses the inherent difficulties in validating MSM. Sec. III.B presents our new approach, which bypasses all these difficulties and allows for validation. Sec. III.C gives a short review of the experimental results used in this work. In in Sec. III.D we compare theory and experiments, and show that MSM is incomplete and requires further evaluation. The implications of which are given in Sec. III.E.

#### A. Difficulties in validating MSM

A direct validation of MSM would require an independent experimental measurement of stresses, which is extremely difficult. In the Stokes formulation, an independent measurement of stress is no longer required, because the stresses follow directly from Eq. (4). Rather, the validation requires that both the tractions, $\boldsymbol{T}$, and velocities, $\boldsymbol{v}$ (these will be denoted by $\boldsymbol{T}^{meas}$ and $v^{meas}$, respectively), be simultaneously measured. Fortunately, both are easily measured: $T^{meas}$ by TFM [7–9] and $v^{meas}$ by PIV, optical flow or other experimental methods.

For the Stokes approach, one end goal would be an analytical solution for the velocity given a traction field, $v^{calc}(T)$, in 2D or 3D, and comparing this with $v^{meas}$. However, this is not without difficulties. First, the governing equation [Eqs. (10) and (14)] are a set of coupled linear partial differential equations. Second, these equations are subject to varying types of BCs (see Sec. II.E). Third, unless the geometry is extremely simple, an analytical solution is typically not possible, thus, necessitating numerical evaluation.

In this work, we suggest an alternative approach that negates the need for solving for $v(T)$. However, we did find two $v(T)$ solutions: Appendix A provides a derivation of a 3D solution, $v_{3D}(T)$, for a monolayer where we assume that the monolayer is infinite in the $x, y$ plane and solve the governing equation via a Fourier Transform. However, due to the complexity of the expressions in $k$-space, the final calculation requires numerical evaluation. Appendix B provides a derivation of the simpler 2D plane-stress problem, which is typically the most relevant for MSM, we calculated these inverse functions and give an exact solution for $v_{2D}(T)$.

### B. Alternative approach to MSM

Once more, the governing equations in 2D are

$$\nabla \cdot v = -p/2\mu, \tag{22}$$

$$-\tfrac{3}{2}\nabla p + \mu \nabla^2 v = T/h. \tag{23}$$

Inserting Eqs. (22) into Eq. (23) leads to [13]

$$\nabla^2 v + 3\nabla(\nabla \cdot v) = T/(\mu h) \equiv t. \tag{24}$$

In classical MSM [11,12,16–25,29], $\sigma_{MSM}(\boldsymbol{T})$, is calculated. In the Stokes formulation, Eq. (24), is solved for $\boldsymbol{v}(\boldsymbol{T})$ [and then $\sigma_{Stokes}(\boldsymbol{T})$ is calculated via Eq. (4)]. As previously stated solving for $\boldsymbol{v}(\boldsymbol{T})$ is non-trivial.

These difficulties can be entirely circumvented. As $\boldsymbol{T}$ and $\boldsymbol{v}$ are both measured, an alternative approach is to solve for $\boldsymbol{T}$ given $\boldsymbol{v}$; denoted as $\boldsymbol{T}^{calc}(\boldsymbol{v})$. Here, we compute the calculated tractions, denoted $\boldsymbol{t}^{calc} = \boldsymbol{T}^{calc}(\boldsymbol{v}^{meas})/\mu h$ [Eq. (24)]. In contrast to $\boldsymbol{v}(\boldsymbol{T})$, which is difficult and BC dependent, calculating $\boldsymbol{t}^{calc}(\boldsymbol{v})$ is simple; it is algebraic, and in-plane edge conditions are not needed. Also, $\sigma_{Stokes}(\boldsymbol{v})$ stresses can be calculated directly (up to a normalization of $\mu$) from the velocities from Eqs. (4) and (22).

We compared $\boldsymbol{t}^{calc}$ with $\boldsymbol{T}^{meas}$ based upon measurements from a previous work wherein $\boldsymbol{v}^{meas}$ and $\boldsymbol{T}^{meas}$ were obtained independently and concurrently in the same experiment [16].

**C.    Experimental setup**

Briefly, cells were cultured until they were confluent in a confined region bounded by a barrier [16]. Subsequently, the barrier was lifted so that the monolayer could migrate into a cell-free space containing a circular obstacle into which the cells cannot migrate. The velocity field was measured by particle image velocimetry (PIV) [9] and the traction field was measured by TFM [8,9]. To smooth the spontaneous spatial fluctuations that are known to characterize collective cellular migration, velocity and traction fields were averaged over six identical experiments, and these fields were smoothed further using a Gaussian filter with a 2 pixel standard deviation [42,43].

The advancing monolayer encountered and then encompassed the circular obstacle (Figure 3a) [16]. The corresponding velocity components $(v_x, v_y)$ around the circular obstacle resembles that of flow around a cylinder [14], in which case the flow field divides at an upstream stagnation point [Figure 3(b), (c)].

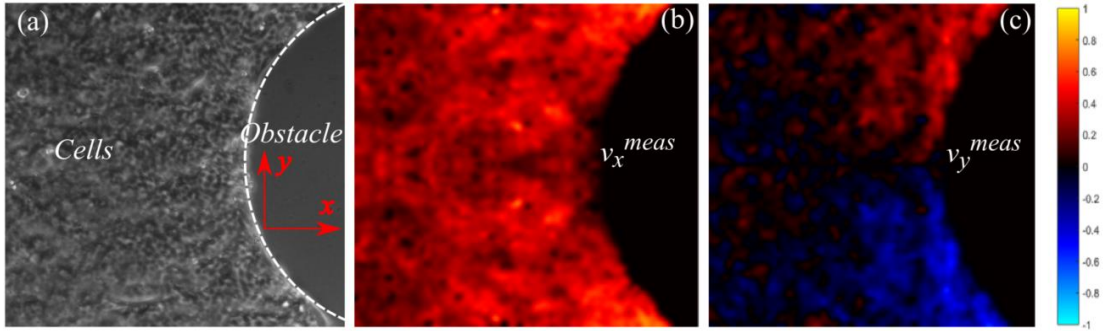

Figure 3. (a) Phase image of the migrating cells encompassing the obstacle. The field of view is 725x725 $\mu m^2$. Measured velocities (b) $v_x$ and (c) $v_y$ (color bar is in units of 10 $[nm/s]$). Raw data from Ref. [16].

**D.    Comparison of $\boldsymbol{T}^{meas}$ and $\mu h \boldsymbol{t}^{calc}$**

After applying the same Gaussian smoothing filter [42,43] to the measured *x*-component of traction $T_x^{meas}$, the heterogeneity and punctate nature of this field became apparent [Figure 4a)]. The corresponding *x*-component of traction, $\mu h t_x^{calc}$, is also heterogeneous, but more highly punctate due to its origin in the derivatives of the velocity field [Figure 4(b)]. The values chosen for the height and viscosity are discussed below (see Sec. IV.A. and Appendix C). A pixel-by-pixel scatter plot of these two traction fields, reveals that the peak of $T_x^{meas}$ is centered around 4 *Pa* while that of $\mu h t_x^{calc}$ is centered around zero (Figure 5). This difference can also be seen in Figure 4(d) and (e) which show the heat maps of $|\boldsymbol{T}^{meas}|$ and $|\mu h \boldsymbol{t}^{calc}|$, respectively, as well as their vector fields. It is evident that $\boldsymbol{T}^{meas}$ [Figure 4(d)] appears to be dominated by the *x*-component, $T_x^{meas}$, and is pulling away from the free edge [9,16], $\mu h \boldsymbol{t}^{calc}$ [Figure 4(e)] appears to lack a distinct direction and is diffuse. To quantify the degree of correlation, we computed the Pearson correlation coefficient between $T_x^{meas}$ and $t_x^{calc}$, given by $\rho_x = \text{cov}(T_x^{meas}, t_x^{calc}) / [std(T_x^{meas}) std(t_x^{calc})]$. This is the covariance of the measured and calculated tractions divided by their respective standard deviations, and is therefore dimensionless and independent of $\mu h$. The calculated correlation coefficients are $\rho_x = -0.022$ and $\rho_y = -0.012$ indicating little correlation (the *y*-component data are given in the Supplementary Material [41]). To ensure that the lack of correlation is not due to the ensemble average or the effect of the interaction with the obstacle we also compared the measured and calculated tractions from a single (non-averaged) dataset pre- and post-interaction with the obstacle [41]. Tractions for these datasets were also uncorrelated.

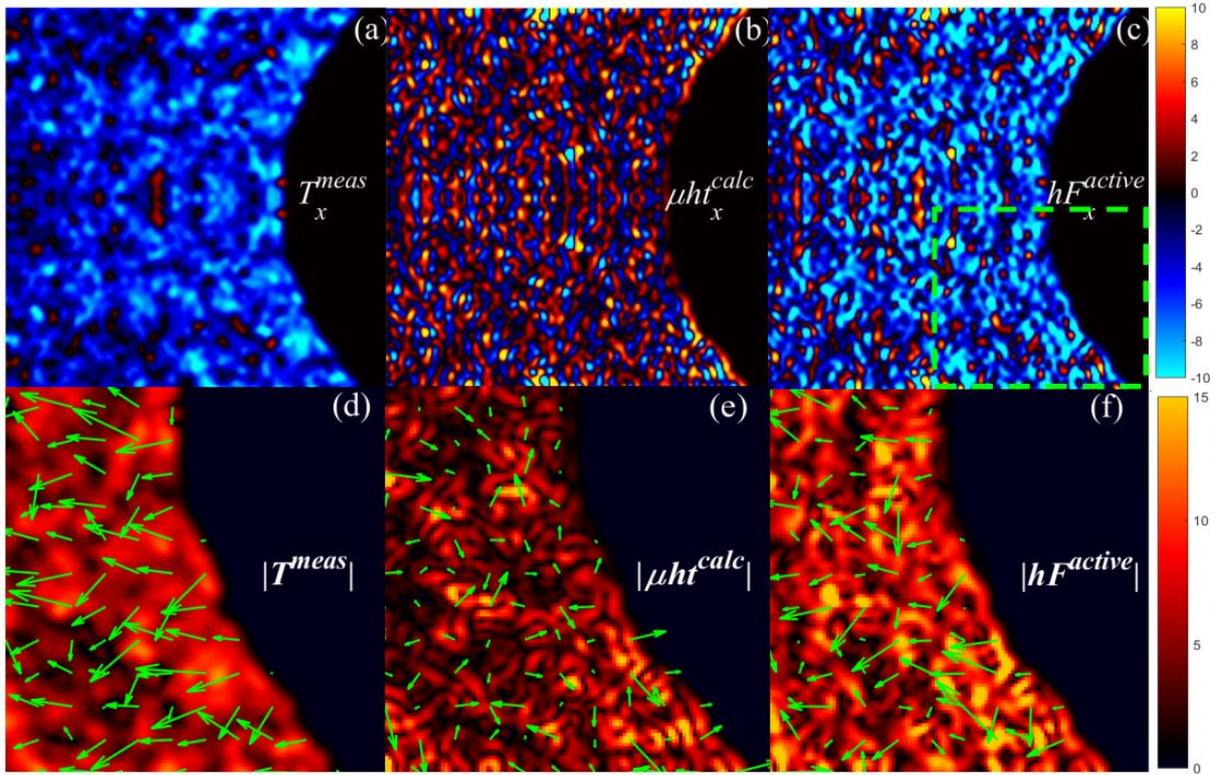

Figure 4.(a) Measured tractions $T_x^{meas}[Pa]$ used in Ref. [16]. (b) Calculated tractions $\mu h t_x^{calc}[Pa]$ with $h = 5\mu m$, and $\mu = 23 kPa \cdot s$. (c) The active force density, $hF_x^{active} = T_x^{meas} - \mu h t_x^{meas}$ [see Eq. (28) below]. (d)-(f) A zoomed view of the marked green box from panel (c), showing (d) $|\boldsymbol{T}^{meas}|[Pa]$, (e) $|\mu h \boldsymbol{t}^{calc}|[Pa]$ and (f) $|h\boldsymbol{F}^{active}|[Pa]$. The green arrows

show the direction of each respective vector of (d)-(f) where the lengths of the arrows are proportional to the magnitude.

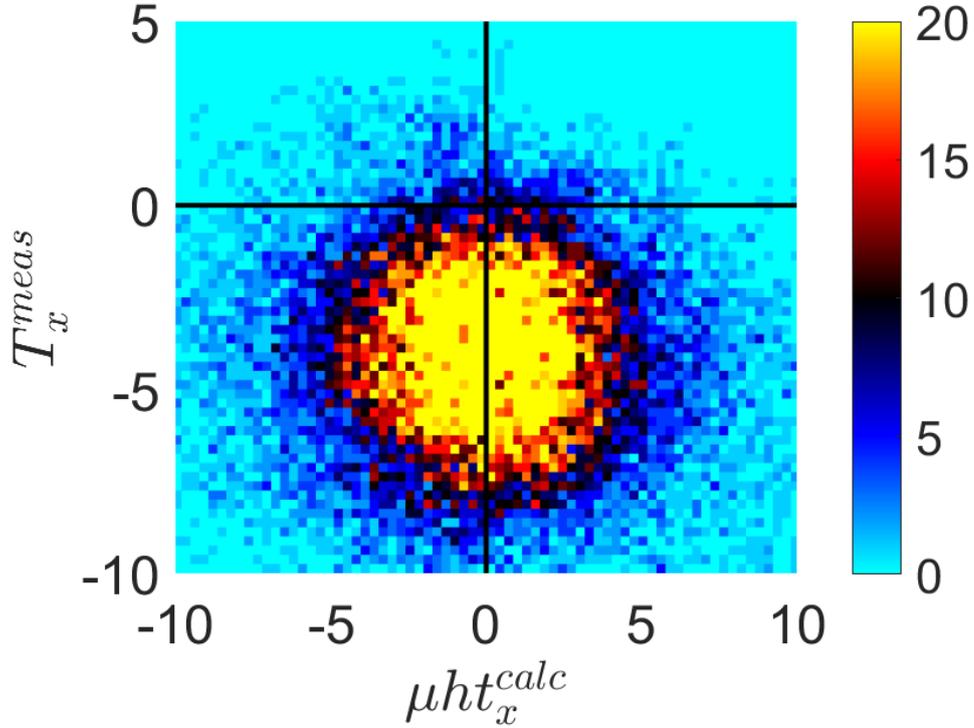

Figure 5. Scatter plot of $T_x^{meas}[Pa]$ versus $\mu h t_x^{calc}[Pa]$. The color bar is proportional to the count density.

### E. Implications for MSM

Until now, the foundations of $\sigma_{MSM}(T)$ have not been verified experimentally because there are no independent measurements of stress. On the other hand, MSM can be tested for internal consistency by a specific comparison of $T^{meas}$ and $T^{calc}(v^{meas})$. Here we have done that. Our finding of a lack of correlation between $T^{meas}$ and $T^{calc}(v^{meas})$ [or $t^{calc}(v)$] necessarily implies $\sigma_{Stokes}(T^{calc}(v^{meas})) \neq \sigma_{MSM}(T^{meas})$. In words, intercellular stresses computed from measured tractions are not the same as intercellular stresses computed from the measured velocity field. This implies that any passive linear formulation, including MSM, is an incomplete formulation.

We point out, that by contrast to the 2D situation, stress recovery using MSM in 1D is straightforward and unambiguous, and a constitutive law is not required. For example, a simple balance of forces requires that stresses in the $x$ direction, $\sigma_{xx} = h^{-1} \int T_x dx$. This approach has been successfully used to calculate 1D stresses [9,17,19,20,26–28,30,31] in cases where the symmetry of the geometry (planar or circular) allows, through averaging, a reduction to 1D.

More complex geometries require solutions in 2D and the need for compatibility equations arising from a constitutive law. For example, Zimmerman et al [21] recently suggested that perhaps stress recovery was method-independent. In that work [21], they compared 2D stresses recovered from 2D tractions using MSM versus the stresses recovered using the molecular dynamic simulation method of Hardy [44]. Heat maps of recovered stresses in that report [21] show close agreement between the two methods at longer length scales, supporting their claim that stress recovery is method-independent, but at shorter length scales indicate discrepancies that approach 400%. At this scale intercellular stress recovery is strongly method-dependent, and as such, the issue of intercellular stress recovery remains open.

## IV. RESOLVING THE MSM PARADOX

The main finding of this work is that tractions calculated from the measured velocity field, $T^{calc}(v^{meas})$, do not correspond to tractions that are measured directly, $T^{meas}$. Hence, the underlying assumptions of MSM require reevaluation. Namely, the constitutive equation must be modified. There are three natural candidates: a linear visco-elastic rheology, a non-linear rheology (visco-elastic or otherwise), or an active rheology.

Linear visco-elastic approaches have been adopted in various biological systems (cell aggregate under aspiration [38], cell aggregate spreading [45], single cell twisting via magnetic tweezers [46,47]. However, to the best of our knowledge, such modeling has not been attempted on monolayers. This approach should be considered in future works on monolayer dynamics. This statement, of future consideration, also holds for the even more difficult non-linear rheologies.

Here, we consider the last of these candidates – active rheology. We do this because currently this is the most favored approach in investigating monolayers dynamics [10,17,19–21,26–31] and there is some experimental evidence supporting this approach. Below, we show a new relation for the behavior of active stresses that can be derived from Newton's law and requires a minimal amount of assumptions.

**A.     Active stresses as a resolution**

Previous works [10,17,19–21,26–31] considered the possibility that the stress tensor additively comprises both a passive, $\sigma^p$, and active, $\sigma^a$, term, such that

$$\boldsymbol{\sigma} = \boldsymbol{\sigma}^p + \boldsymbol{\sigma}^a. \tag{25}$$

It is then assumed that [10,17,19–21,26–31], the passive term is either an elastic solid [Eq. (2)] or viscous fluid [Eq. (4)]. To the extent that the velocities are representative of the motion of the composite monolayer, we insert $\sigma^p = \sigma^{viscous}$ and Eq. (25) into Newton's law [Eq.(11)]

$$\nabla \cdot (\boldsymbol{\sigma}^a + \boldsymbol{\sigma}^p) = \boldsymbol{T}/h. \tag{26}$$

The active force density, denoted $\boldsymbol{F}^{active}$, is defined as

$$\boldsymbol{F}^{active} = \nabla \cdot \boldsymbol{\sigma}^a. \tag{27}$$

From Eqs. (4),(26) and (27) we have $\boldsymbol{F}^{active}$ explicitly in terms of measured quantities:

$$\begin{aligned} h\boldsymbol{F}^{active} &= \boldsymbol{T}^{meas} - \mu h \boldsymbol{t}^{calc} \\ &= \boldsymbol{T}^{meas} - \mu h [\nabla^2 \boldsymbol{v}^{meas} + 3\nabla(\nabla \cdot \boldsymbol{v}^{meas})] \end{aligned} \tag{28}$$

$\boldsymbol{F}^{active}$ includes not only the tractions and velocities, but also the height $h$ and the viscosity $\mu$. It follows from Eq. (28) that in the limit of low viscosity $\boldsymbol{F}^{active}$ is dominated by the tractions $\boldsymbol{T}^{meas}$, whereas in the limit of high viscosity $\boldsymbol{F}^{active}$ is dominated by passive component $\mu h \boldsymbol{t}^{meas}$.

Note that while $h$ may be estimated accurately [9,10,18], the same is not true for $\mu$. Appendix C shows a number of estimates using different methods [41]. Here we illustrate $F_x^{active}$ for the case where $h = 5\mu m$ and $\mu = 23 kPa \cdot s$, the latter being our best estimate resting on the fewest assumptions, and also falling in an intermediate range such that the contributions of both $T^{meas}$ and $\mu h t^{meas}$ can both be seen [Figure 4(c)]. For these values, and for the data set described above, we used Eq. (28) to compute $hF_x^{active}$ [Figure 4(c)], which shows clear contributions from both the traction term [Figure 4(a)] and the passive velocity terms [Figure 4(b)]. For example, it can be observed that $hF_x^{active}$ is similar to $T_x^{meas}$ in that the tractions at the free edge are largely pointing away from the obstacle [Figure 4(a)], as was previously shown [9]. Similarly, there is fine scale structure in $hF_x^{active}$ which reflects its origin in the passive component [Figure 4(b)]. Figure 4(f) is a vector map showing the directions $\boldsymbol{F}^{active}$ where the length of the arrows are proportional to $|\boldsymbol{F}^{active}|$. Note that the active forces are largely pointing away from the obstacle, similar to the tractions, as was shown in [9,16].

### B. Origin of active stresses

Recent experiments have confirmed the existence of active forces in epithelial monolayers. For example, in Ref. [26] it was found that stresses are linearly associated with strains but with a nonzero offset, expressed as $\sigma = E\varepsilon + \sigma^a$. This is interpreted as reflecting an active component in both the modulus and the offset [26] and supports the active rheology suggested by Eq. (25). Two recent works also used Eq. (25) in their derivations to explain wetting [31] and fingering instabilities [30] observed in epithelial monolayers. However, they didn't consider the question of whether or not that the passive formulation is incomplete.

With respect to the origin of active forces, velocities, and tractions, several previous works have suggested possible phenomenological forms for the active stress [10]. These potential origins of $F^{active}$ differ substantially in the literature. Some works suggest the origin is in the concentration of active contractile units [19,20] while other suggests in-plane cell polarization [19,20,29–31,48], non-zero tension [26], cell division [26,35]. Importantly, without a compatibility equation for $\sigma^a$, integrating Eq. (28) to calculate $\sigma^a$, and thus the total stress, $\sigma$, is not possible. The current state of knowledge regarding $F^{active}$ is inconclusive as to which of the above models is most appropriate [10]. Nevertheless, Eq. (28) describing the active stress is derived from Newton's laws, with a minimal number of embedded assumptions, and must be satisfied.

## V.  CONCLUSIONS

We have shown that, based on a linear passive constitutive law (either elastic or viscous), tractions computed from the measured velocity field are uncorrelated with measured tractions. This implies that MSM as a method is incomplete and requires further evaluation. We suggest that intercellular active stresses may resolve this inconsistency, but that if the active stresses are not derivable from a constitutive law, then it may not be possible to explicitly compute intercellular stresses. However, we do give an explicit expression [Eq.(28)] for the active force density as a function of the measured tractions and velocities. Regardless of the proposed origin of the active stresses, this condition, derived from Newton's 2nd law, must be satisfied.

## ACKNOWLEDGMENTS

We thank Dr. Jae Hun Kim for generously providing us with the raw experimental data used in Ref. [16], and used in this work. We also thank Dr. Bo Lan for helpful discussions. This work was supported by National Institutes of Health (NIH) Grants U01CA202123, PO1HL120839 and T32HL007118.

# APPENDIX A: 3D SOLUTION FOR $v(T)$

## A. Governing equations in real space

In 3D, the governing Stokes equations for an incompressible fluid of uniform density are

$$\nabla \cdot \boldsymbol{v} = 0, \tag{29}$$

$$-\nabla p + \mu \nabla^2 \boldsymbol{v} = 0, \tag{30}$$

where $\boldsymbol{v} = (v_x, v_y, v_z)^T$ is the 3D velocity vector in Cartesian coordinates and $p$ is the pressure [Figure 1 (a)]. Equations (29) and (30) express conservation of mass and force balance, respectively.

For boundary conditions, we take the top (apical) surface, $z = h$, to be stress free ($\boldsymbol{\sigma} \cdot \boldsymbol{n} = 0$) [Figure 1(b)]:

$$\mu(v_{x,z} + v_{z,x})\big|_{z=h} = 0, \tag{31}$$

$$\mu(v_{y,z} + v_{z,y})\big|_{z=h} = 0, \tag{32}$$

$$(2\mu v_{z,z} - p)\big|_{z=h} = 0. \tag{33}$$

The comma subscript denotes partial differentiation. Note that we assume the film height to change only slowly ($\nabla h \ll 1$) such that curvatures terms are negligible [49–51]. At the bottom (basal) surface that is in contact with the substrate, $z = 0$, the shear stress must be balanced by the tractions, and we take zero normal flux:

$$v_{x,z}\big|_{z=0} = T_x / \mu, \tag{34}$$

$$v_{y,z}\big|_{z=0} = T_y / \mu, \tag{35}$$

$$v_z\big|_{z=0} = 0. \tag{36}$$

## B. Governing equations in $k$ space

The form of Eqs. (29)-(36) suggests using 2 dimensional Fourier transforms, defined for any function $f(\mathbf{r})$ by $\tilde{f}(\mathbf{k}) = \int d^2 r f(\mathbf{r}) e^{-i\mathbf{k}\cdot\mathbf{r}}$, integrated over the plane. Here $\mathbf{r} = (x, y)^T$, the wave vector $\mathbf{k} = (\alpha, \beta)^T$, $k = |\mathbf{k}|$, $r = |\mathbf{r}|$ and $i = \sqrt{-1}$ is the imaginary unit. With this notation, Eqs. (29)-(36) transform to

$$i\alpha\tilde{v}_x + i\beta\tilde{v}_y + \tilde{v}_{z,z} = 0, \tag{37}$$

$$-i\alpha\tilde{p} + \mu(\tilde{v}_{x,zz} - k^2\tilde{v}_x) = 0, \tag{38}$$

$$-i\beta\tilde{p} + \mu(\tilde{v}_{y,zz} - k^2\tilde{v}_y) = 0, \tag{39}$$

$$-\tilde{p}_{,z} + \mu(\tilde{v}_{z,zz} - k^2\tilde{v}_z) = 0. \tag{40}$$

Note that this approach to solving the field equations is exactly the same as that used in Ref. [9], albeit with different top and bottom boundary conditions. In the current application, the transformed stress-free boundary conditions are given by, for the top surface

$$(\tilde{v}_{x,z} + i\alpha\tilde{v}_z)\big|_{z=h} = 0, \tag{41}$$

$$(\tilde{v}_{y,z} + i\beta\tilde{v}_z)\big|_{z=h} = 0, \tag{42}$$

$$(2\mu\tilde{v}_{z,z} - \tilde{p})\big|_{z=h} = 0, \tag{43}$$

and for the tractions and homogeneous flux at the bottom

$$\tilde{v}_{x,z}\big|_{z=0} = \tilde{T}_x / \mu, \tag{44}$$

$$\tilde{v}_{y,z}\big|_{z=0} = \tilde{T}_y / \mu, \tag{45}$$

$$\tilde{v}_z\big|_{z=0} = 0. \tag{46}$$

## C.  3D Solution in *k* space

There are 6 solutions to Eqs. (37)-(40); these are given in Ref. [9]. The two linear combinations of which satisfy the four homogeneous conditions [Eqs. (41)-(43) and (46)] can be written in vector notation

$$\varphi_1(z) = \frac{\cosh[k(h-z)]}{\cosh kh} \begin{bmatrix} \beta \\ -\alpha \\ 0 \end{bmatrix}, \tag{47}$$

$$\varphi_2(z) = -\frac{1+2k^2h^2+\cosh 2hk}{\phi} \begin{bmatrix} \alpha \cosh kz \\ \beta \cosh kz \\ -ik \sinh kz \end{bmatrix} + \begin{bmatrix} \alpha \sinh kz \\ \beta \sinh kz \\ -ik \cosh kz \end{bmatrix} + kz \begin{bmatrix} \alpha \cosh kz \\ \beta \cosh kz \\ -ik \sinh kz \end{bmatrix}$$
$$-\frac{2(\cosh hk)^2}{\phi} kz \begin{bmatrix} \alpha \sinh kz \\ \beta \sinh kz \\ -ik \cosh kz \end{bmatrix} + \begin{bmatrix} 0 \\ 0 \\ ik(\cosh kz - 2\phi^{-1}(\cosh hk)^2 \sinh kz) \end{bmatrix}, \tag{48}$$

where $\phi = 2hk + \sinh(2hk)$. The corresponding solutions for the pressure are

$$p_1 = 0, \quad p_2 = -2ik^2 \sinh kz + \frac{4ik^2(\cosh hk)^2}{\phi} \cosh kz. \tag{49}$$

In these terms, write the linear combination as

$$\mathbf{v} = \begin{bmatrix} \tilde{v}_x \\ \tilde{v}_y \\ \tilde{v}_z \end{bmatrix} = A\varphi_1 + B\varphi_2, \quad p = A\mu p_1 + B\mu p_2, \tag{50}$$

where $A, B$ are chosen to satisfy the inhomogeneous traction conditions [Eqs. (44)-(45)]. Explicitly, this requires

$$M \begin{bmatrix} A \\ B \end{bmatrix} = \begin{bmatrix} \tilde{T}_x / \mu \\ \tilde{T}_y / \mu \end{bmatrix}, \tag{51}$$

where

$$M = k \begin{bmatrix} -\beta \tanh kh & 2\alpha \\ \alpha \tanh kh & 2\beta \end{bmatrix}. \tag{52}$$

This leads to

$$\begin{bmatrix} A \\ B \end{bmatrix} = M^{-1} \begin{bmatrix} \tilde{T}_x/\mu \\ \tilde{T}_y/\mu \end{bmatrix}. \quad (53)$$

Specifically,

$$A = \frac{\alpha T_y - \beta T_x}{\mu k^3} \coth kh, \quad B = \frac{\alpha T_x + \beta T_y}{2\mu k^3}. \quad (54)$$

Note that in experiments with transverse averaging (such that $v_y = 0$), and where the height is spatially uniform, there is a simple volume conservation relation between $v_z(z=h)$ and the height averaged $v_x$ at the two edges, say $x = a$ and $x = b$. Explicitly,

$$h[v_x(x=b) - v_x(x=a)] + \int_a^b v_z(x'; z=h) dx' = 0. \quad (55)$$

In 2D expanding islands, the equivalent expression of volume conservation is

$$h \oint_C ds \, \mathbf{n} \cdot \mathbf{v} + \int_D dx dy \, v_z(x,y; z=h) = 0, \quad (56)$$

where $C$ and $D$ are the boundary and domain, respectively. If proliferation is present, these expressions give the volume averaged mean proliferation rate, given measurements of both edge and height velocities.

**APPENDIX B: 2D SOLUTION FOR $v(T)$**

**A. 2D solution in $k$ space**

The 2D governing equations are

$$\nabla \cdot \mathbf{v} = -p/2\mu, \quad (57)$$

$$-\tfrac{3}{2}\nabla p + \mu \nabla^2 \mathbf{v} = \mathbf{T}/h. \quad (58)$$

Solution of Eqs. (57)-(58) for the case of infinite and periodic edge BCs (similar to Appendix A) yields

$$\tilde{p} = \tilde{\mathbf{G}}_p(\mathbf{k}) \cdot \tilde{\mathbf{T}}, \quad (59)$$

$$\tilde{G}_p(k) = \frac{i\mathbf{k}}{2hk^2}, \tag{60}$$

$$\mu\tilde{\mathbf{v}} = \frac{3(\mathbf{k}\cdot\tilde{\mathbf{T}})\mathbf{k}}{4hk^4} - \frac{\tilde{\mathbf{T}}}{hk^2}. \tag{61}$$

With tractions as the only input, inserting Eq. (59)-(61) into the constitutive law [Eq. (4)] confirms the scaling argument in Refs. [11,12] that the stresses are independent of the viscosity (or shear modulus).

**B.    2D solution in real space**

We can now calculate Green's function in real space, i.e. the impulse response corresponding to a delta function of tractions at the origin. This is given by the inverse Fourier transform $f(\mathbf{r}) = (4\pi^2)^{-1}\int d^2k \tilde{f}(\mathbf{k})e^{i\mathbf{k}\cdot\mathbf{r}}$. We start with the pressure [Eq. (59)],

$$\begin{aligned}\mathbf{G}_p(\mathbf{r}) &= \frac{1}{4\pi^2}\int_{-\infty}^{\infty} d^2k\,\tilde{\mathbf{G}}_p(\mathbf{k})e^{i\mathbf{k}\cdot\mathbf{r}} = \frac{1}{4\pi^2}\int_{-\infty}^{\infty} d^2k\,\frac{i\mathbf{k}}{2hk^2}e^{i\mathbf{k}\cdot\mathbf{r}} = \frac{1}{4\pi^2}\int_{-\infty}^{\infty} d^2k\,\frac{\nabla e^{i\mathbf{k}\cdot\mathbf{r}}}{2hk^2}\\ &= \frac{1}{8\pi^2 h}\int_0^{\infty} dk\,\frac{1}{k}\nabla\int_0^{2\pi} d\theta\,e^{ikr\cos\theta} = \frac{1}{8\pi^2 h}\int_0^{\infty} dk\,\frac{2\pi}{k}\nabla J_0(kr) = -\frac{\nabla r}{4\pi h}\int_0^{\infty} dk J_1(kr) = -\frac{\mathbf{r}}{4\pi h r^2}\end{aligned} \tag{62}$$

Thus, the pressure is given by the convolution, integrated over all space,

$$p = -\int d^2 r'\,\frac{\tilde{\mathbf{T}}(\mathbf{r}')\cdot(\mathbf{r}-\mathbf{r}')}{4\pi h |\mathbf{r}-\mathbf{r}'|^2}. \tag{63}$$

For the velocity, we identify that we need to calculate three different terms $\alpha^2 k^{-4}, \beta^2 k^{-4}, \alpha\beta k^{-4}$ [$\mathbf{k}=(\alpha,\beta)^T$]. The first term is

$$\begin{aligned}FT^{-1}\left(\frac{\alpha^2}{k^4}\right) &= \frac{1}{4\pi^2}\int_{-\infty}^{\infty} d^2k\,\frac{\alpha^2}{k^4}e^{i\mathbf{k}\cdot\mathbf{r}} = \frac{1}{4\pi^2}\int_{-\infty}^{\infty} d^2k\,\frac{-\alpha}{2}\left[\frac{\partial}{\partial\alpha}\left(\frac{1}{k^2}\right)\right]e^{i\mathbf{k}\cdot\mathbf{r}}\\ &= \frac{i}{8\pi^2}\frac{\partial}{\partial x}\int_{-\infty}^{\infty} d^2k\left[\frac{\partial}{\partial\alpha}\left(\frac{1}{k^2}\right)\right]e^{i\mathbf{k}\cdot\mathbf{r}} = -\frac{1}{8\pi^2}i\frac{\partial}{\partial x}\left[(ix)\int_{-\infty}^{\infty} d^2k\,\frac{1}{k^2}e^{i\mathbf{k}\cdot\mathbf{r}}\right]\end{aligned}, \tag{64}$$

To find $\int_{-\infty}^{\infty} d^2k \, k^{-2} e^{ik \cdot r}$, consider the impulse response to the 2D Laplace equation, $\nabla^2 G = \delta^2(r)$, for which the solution is $(2\pi)^{-1} \ln r$. Its 2D Fourier transform is $\tilde{G} = -k^{-2}$. This implies that $\int_{-\infty}^{\infty} d^2k \, k^{-2} e^{ik \cdot r} = -(2\pi)^{-1} \ln r$ from which it is immediate that

$$FT^{-1}\left(\frac{\alpha^2}{k^4}\right) = -\frac{1}{8\pi^2} \frac{\partial}{\partial x}\left[(2\pi)^{-1} x \ln r\right] = -\frac{1}{16\pi^3}\left(\frac{x^2}{r^2} + \ln r\right), \tag{65}$$

$$FT^{-1}\left(\frac{\beta^2}{k^4}\right) = -\frac{1}{16\pi^3}\left(\frac{y^2}{r^2} + \ln r\right), \tag{66}$$

$$FT^{-1}\left(\frac{\alpha\beta}{k^4}\right) = -\frac{1}{16\pi^3}\left(\frac{xy}{r^2} + \ln r\right). \tag{67}$$

**APPENDIX C: VISCOSITY ESTIMATES**

Despite the apparent inconsistencies of using only a passive stress tensor, we use the pure fluid dynamical description to estimate the monolayers viscosity. We employ two different independent methods. First, consider the scaling argument that viscous stresses in a thin film balance the tractions, $\mu \sim Th/u$. For typical values of $T \sim 10 [Pa]$, $u \sim 10 [nm/s]$, and $h \sim 5 [\mu m]$ [9,11,17] one gets $\mu_{cell} \sim 5 [kPa \cdot s] \sim 5 \cdot 10^6 \mu_{water}$ with $\mu_{water} = 10^{-3} [Pa \cdot s]$. Next, we considered a simple coarse grained calculation $\mu = \langle T^{meas}\rangle_{rms} / \langle |t^{calc}| h\rangle_{rms}$ where $\langle \, \rangle_{rms}$ is the root-mean-square average; this gives the result $\mu/\mu_{water} \approx O(2.3 \cdot 10^7)$. These values, together with others from the literature are shown in Table 1. Our values are on the lower end of the reported range, but we note that all these estimates are extremely large compared with what we consider physically reasonable. This is another indication that a purely passive viscous description is inadequate, and points to the necessity of active stresses to account for the relationship between velocities and tractions.

Table 1. Viscosity estimates from various experimental setups. The first two rows are from this work.

| *Method* [References] | $10^6 \times (\mu / \mu_{water})$ |
|---|---|
| *Scaling* ($Th/u$) | 5 |
| *Coarse graining* ($\langle T^{meas} \rangle_{rms} / \langle |t^{calc}|h \rangle_{rms}$) | 23 |
| *Low frequency* $G''$ [46] | 1 |
| *Micropipette aspiration* [38] | 200 |
| *Expanding island assays* [29] | $10^2$-$10^4$ |
| *Expanding island – active stresses* [30,31] | $(0.5$-$5) \times 10^3$ |